\documentclass[aip,jap,amsmath,amssymb,
reprint, footinbib,unsortedaddress]{revtex4-1}
\usepackage{CJK}
\usepackage{dcolumn}
\usepackage{graphicx}
\usepackage{amssymb}
\usepackage{bm}
\usepackage{dcolumn}
\usepackage{bm}

\begin{document}
\begin{CJK*}{SJIS}{MS Mincho} 

\title{Signatures of d-Wave Symmetry on Thermal Dirac Fermions in Graphene-Based
F/I/d Junctions}

\author{Morteza Salehi}\thanks{\textbf{Current address}: \textit{Department of Physics, Sharif University of Technology, Tehran 11155-9161, Iran.}}

\author{Mohammad Alidoust}\thanks{\textbf{Current address}: \textit{Department of Physics, Norwegian University of Science and
Technology, N-7491 Trondheim, Norway.}}

\author{Gholamreza Rashedi}

\affiliation{Department of Physics, Faculty of Sciences, University of
Isfahan, Hezar Jerib Avenue, Isfahan 81746-73441, Iran}

\date{\today}
\begin{abstract}
We study theoretically the behavior of thermal massless Dirac
fermions inside graphene-based Ferromagnetic $\mid$ Insulator $\mid$
 $d$-wave/$s$-wave superconductor (F$\mid$I$\mid$d and
F$\mid$I$\mid$S ) junctions in the ballistic regime. Using the
Dirac-BdG wave functions within the three regions and appropriate
boundary conditions, the Andreev and Normal reflection coefficients
are derived. By employing the obtained Andreev and Normal reflection
coefficients the characteristics of heat current through the
F$\mid$I$\mid$d and F$\mid$I$\mid$S junctions are investigated
within the thin barrier approximation. We find that for $s$-wave
superconductors, thermal conductance oscillates sinusoidally vs
barrier strength. The finding persist for the values of $\alpha$,
the orientation of $d$-wave superconductor crystal in the $k$-space,
below $\pi/4$. By increasing temperature, the thermal conductance is
increased exponentially for small values of $\alpha$
 and for larger values the quantity is modified to exhibit a linear behavior at
$\alpha=\pi/4$ which is similar to Wiedemann-Franz law for metals in
low temperatures.
\end{abstract}
\maketitle

\end{CJK*}

\section{\label{sec:intro}Introduction}
Graphene is a single layer of carbon atoms which was introduced to
the scientific community by Novoselov \textit{et al.} in $2004$
\cite{novoselov1,novoselov2}. Most of applicable and interesting
characteristics of graphene has been investigated intensely
experimentally
\cite{novoselov1,novoselov2,novoselov3,novoselov4,novoselov5,novoselov6}.
Because of interesting phenomena which graphene showed, the
artificial material has been received robust attentions from
theoretical and experimental physics communities
\cite{beenakker2,beenakker3,beenakker1,beenakker4}. Induction of
superconductive correlations into
 graphene layer by proximity of superconducting electrodes observed
experimentally by Heersche \textit{et al.}\cite{Heersche}. Also
inducing the ferromagnetism into graphene layers by means of
proximity effects observed experimentally by Tombros \textit{et
al.} \cite{Tombros}. In this regard, theoretical scientists
utilized Dirac-Bogoliubov-de Genne (Dirac-BdG) for investigating
and predicting interesting phenomena because of the proximity
effects
\cite{beenakker2,beenakker1,Blonder-Tinkham-klapwijk,linder1,zareyan1,Cayssol}.

Interplay between ferromagnetic graphene sheets and conventional
superconductor is generalized theoretically \cite{zareyan1}. Also
Linder \textit{et al.} \cite{linder2} generalized the theoretical
investigations from conventional to unconventional superconductors
in similarity with metallic cases which has been studied
intensely\cite{tanaka1,tanaka2,hu} . Most of the previous works are
devoted to study electronic transport properties of the
graphene-based junctions such as; Josephson
currents\cite{beenakker3,asano}, electronic
conductance\cite{beenakker1,Sengupta}, spintronic
conductance\cite{salehi1,Hsu}, shot noise \cite{beenakker4,sonin}
and etc., but poor attentions has been focused on the heat transport
properties and electronic thermal conductance of the junctions. The
BTK formalism is generalized by Bardas and Averin \cite{bardas} for
obtaining electronic thermal conductance in the clean limit. Also,
Devyatov \textit{et al.} \cite{devyatov1,devyatov2} studied
electronic thermal conductance of Normal metal$\mid$I$\mid$d
junctions in the ballistic regime. For high-sensitive devices
including graphene-based junctions, knowing all electronic and
thermal properties of the junctions are crucial and important points
from application point of view. Few previous works devoted for
investigating of electronic thermal transport characteristics of the
graphene-based junctions \cite{linder3,salehi2}.
\par
In this paper we especially investigate signatures of
$d_{x^2-y^2}$-wave symmetry on the electronic heat transport
characteristics of the F$\mid$I$\mid$d junctions. We start with the
Dirac-BdG Hamiltonian and use obtained wave functions within the
three regions and appropriate boundary conditions at interfaces for
deriving the Andreev and Normal reflection coefficients in the thin
barrier approximation \cite{Sengupta,Sengupta2}. Using the mentioned
coefficients we present numerical investigations of electronic
thermal conductance of the
Ferromagnetic$\mid$Insulator$\mid$$d$/$s$-wave superconductor
junctions in the ballistic limit. We find that for F$\mid$I$\mid$d
junctions the electronic thermal conductance, $\Gamma$ shows
oscillatory behavior vs barrier strength. Increasing the orientation
angle of the superconducting gap $\alpha$ up to values close to
$\pi/4$ only enhances whole values of the $\Gamma$ and for maximum
value of the superconducting gap orientation angle namely
$\alpha=\pi/4$ magnitude of the oscillations diminish highly.
Another finding is that the electronic thermal conductance shows an
exponential increase vs temperature for small values of $\alpha$. By
increasing the angle of superconducting gap orientation up to
$\pi/4$ the mentioned exponential form modify to linear form and
\textit{at $\alpha=\pi/4$ the thermal conductance shows precisely
linear increase with respect to temperature, namely $\Gamma\propto
T$ that the finding induces in mind the Wiedemann-Franz law
\cite{wiedemann} from metals in the low temperatures}. The paper is
organized as follows:
\par
In Sec. \ref{theory} we explain the analytical derivations of
Andreev and Normal reflection coefficients by starting from the
Dirac-BdG Hamiltonian and in Sec. \ref{FIS} the electronic thermal
conductance of F$\mid$I$\mid$S is investigated by plotting the
quantity with respect to strengthes of exchange field and barrier in
the thin barrier approximation regime. In Sec. \ref{FId} we study
the effects of d$_{x^2-y^2}$-wave symmetry on the electronic thermal
conductance of the junctions within the thin barrier approximation.
The paper go to end with conclusions in Sec. \ref{summary}.

\section{\label{theory}Theory}

We study interplay between graphene-based ferromagnetic and
superconductor junctions in the ballistic limit, therefore we employ
the Dirac-BdG Hamiltonian for obtaining suitable wave functions. The
general Dirac-BdG equation incorporating ferromagnetism and
superconductivity reads as \cite{zareyan1}:
\begin{eqnarray}
\left(
\begin{array}{ccc}
 \nonumber H_0-\sigma h & \Delta(T) \\
  \Delta^{*}(T) & -(H_0-\bar{\sigma} h)
\end{array}
\right)\left( \begin{array}{ccc}
  u_\sigma \\
  v_{\bar{\sigma}}
\end{array}\right)=\epsilon_\sigma\left( \begin{array}{ccc}
  u_\sigma \\
  v_{\bar{\sigma}}
\end{array}\right), \label{BdG}
\end{eqnarray}
\begin{equation}H_{0}({\bf r})=-i\hbar
v_{F}(\sigma_x\partial_x+\sigma_y\partial_y)+U({\bf
r})-E_{F}
\end{equation}
where $\sigma_x$ and $\sigma_y$ are $2 \times 2$ Pauli matrices and
$\Delta(T)$ stands for temperature-dependent order parameter of
superconducting region, also $h$ represents the strength of exchange
field in the ferromagnetic region. $\epsilon_{\sigma}$ stands for
excitation energies of holelike and electronlike quasiparticles. For
obtaining Dirac-BdG wave functions in the Normal, ferromagnetic and
superconducting region, one should set $\{h=0, \Delta(T)=0\}$,
$\{h\neq 0, \Delta(T)=0\}$ and $\{h=0, \Delta\neq 0\}$ inside Eq.
(\ref{BdG}), respectively. Here $\sigma=\pm1$ stands for spin-up and
-down quasiparticles and $\bar{\sigma}=-\sigma$. Also $U({\bf r})$
shows the Fermi energy mismatch. Throughout the paper we consider a
step function for spatial-dependency of the superconducting gap,
namely $\Delta(x,T)=\Delta(T)\Theta(x)$ in which $\Theta(x)$ is the
well known step function. By solving the Eq. (\ref{BdG}) in the
Ferromagnetic region we obtain Dirac-BdG wave functions for
electronlike and holelike quasiparticles as follows:

\begin{align}
 \left\{\begin{array}{cc}
\psi_{e,\sigma}^{\pm}(x)=\frac{1}{\sqrt{\cos{\theta_\sigma}}}\left(
     1,
     \pm e^{\pm i \theta_\sigma},
     0,
     0
   \right)^T e^{(\pm
ik_{e,\sigma} x)}\\
\psi_{h,\bar{\sigma}}^{\pm}(x)=\frac{1}{\sqrt{\cos{\theta_{\bar{\sigma}}}}}\left(
     0,
     0,
     1,
     \mp e^{\pm i \theta^{\prime}_{\bar{\sigma}}}
   \right)^T e^{(\pm
ik_{h,\bar{\sigma}} x)} ,\end{array}
    \right.
\end{align}
where $\theta_\sigma({\theta^{\prime}}_{\bar{\sigma}})$ are
propagation angles of electronlike (holelike) quasiparticles with
respect to the normal trajectory into the interface at $x=0$. We
define the two incident angles as
\begin{eqnarray}\label{incident angle}
 \left\{\begin{array}{cc}
\theta_{\sigma}=\arcsin{\left(\frac{\hbar v_F q}{\epsilon+E_F+\sigma
  h}\right)}\\
 {\theta^{\prime}}_{\bar{\sigma}}=\arcsin{\left(\frac{\hbar v_F q}{\epsilon-E_F+\sigma
    h}\right)},
\end{array}
    \right.
\end{eqnarray}
and $x$-components of the wave vectors for electronlike and holelike
quasiparticles in the Ferromagnetic region are obtain as
 \begin{align}
 \left\{\begin{array}{cc}
k_{e,\sigma}=\frac{\epsilon+E_F+\sigma h}{\hbar
 v_F}\cos\theta_{\sigma}\\
k_{h,\bar{\sigma}}
   =\frac{\epsilon-E_F+\sigma h}{\hbar
   v_F}\cos\theta^{\prime}_{\bar{\sigma}},
\end{array}
    \right.
\end{align}
\begin{figure}
\includegraphics[width=8.5cm,height=3cm]{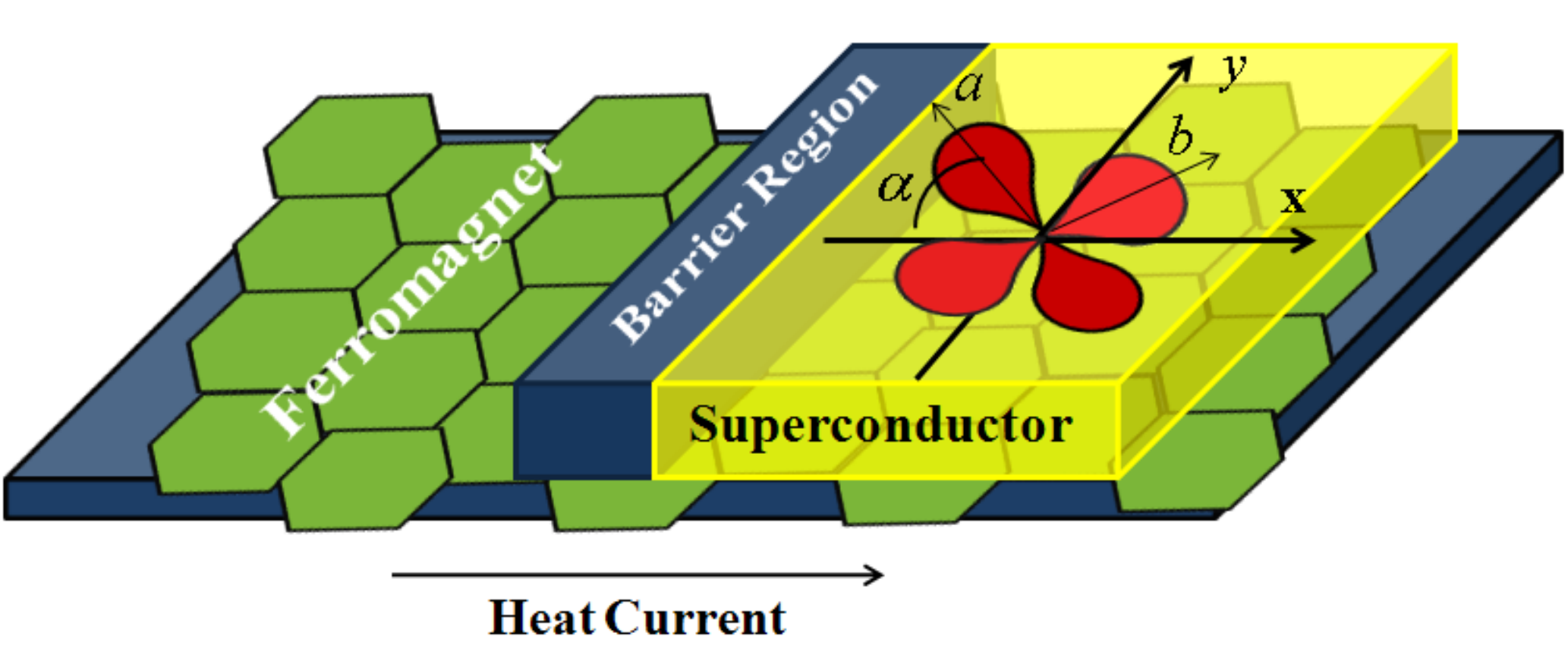}
\caption{\label{fig:model} (Color online) Model for suggested set up
of Ferromagnetic$\mid$Barrier$\mid$$d$-wave superconductor
graphene-based junctions. The junction is located at $x=0$. Axes of
$d$-wave superconductor crystal are represented with $a$ and $b$.
The superconductor crystal orientation angle with respect to normal
trajectory ($x$-axis) is shown by $\alpha$.}
\end{figure}
in which $E_F$ and $q$ are Fermi energy and $y$-component of wave
vector in the Ferromagnetic region, respectively. For normal region
with a barrier potential $V_0$, it is sufficient to set $h=0$  and
$E_F\rightarrow (E_F-V_0)$ in the above obtained equations inside
the Ferromagnetic region. The barrier potential $V_0$ can be applied
by a gate voltage into the region. Within the $d$-wave
superconductor region ($x>0$), Dirac-BdG wave functions for
electronlike and holelike quasiparticles are obtained:
\begin{align}
 \left\{\begin{array}{cc}
\psi^{+}_{S,e}=\left(
     e^{i\beta_{+}},
     e^{i\beta_{+}+i \gamma_{+}},
     e^{-i\phi_{+}},
     e^{i\gamma_{+}-i\phi_{+}}
   \right)^T e^{-i(k_0-i\chi_+) x}\\
\psi^{-}_{S,h}=\left(
     e^{-i\beta_{-}},
     -e^{-i\beta_{-}-\gamma_{-}},
     e^{i\phi_{-}},
     -e^{-i\gamma_{-}-i\phi_{-}}
   \right)^T \\e^{ -i(k_0-i\chi_-) x}
\end{array}
    \right.\label{wf_h}
\end{align}
where we define $\chi_{\pm}=(U_0+E_F)\sin{\beta_{\pm}}/{k_0(\hbar
v_F)^2}$ in which $k_0$ is defined as $k_{0}=((\frac{U_0+E_F}{\hbar
v_F})^2-q^2)^{1/2}$. In Eq. (\ref{wf_h}) $e^{i\beta_{\pm}}$ is
defined as $u_\pm /v_\pm$ and

\begin{eqnarray}
 \left\{\begin{array}{cc}
u_\pm
=\sqrt{\frac{1}{2}(1+\frac{\sqrt{\epsilon^2-|\Delta(\gamma_\pm)}|^2}{\epsilon})}\\
 v_\pm
=\sqrt{\frac{1}{2}(1-\frac{\sqrt{\epsilon^2-|\Delta(\gamma_\pm)}|^2}{\epsilon})}
\end{array}
    \right.
\end{eqnarray}
\begin{equation}\label{beta}
\beta_{\pm}=\left(\begin{array}{c}
     \cos^{-1}(\frac{\epsilon}{\Delta(\gamma_\pm)}),\qquad \epsilon<\Delta(\gamma_\pm) \\
     -i\cosh^{-1}(\frac{\epsilon}{\Delta(\gamma_\pm)}),\qquad \epsilon>\Delta(\gamma_\pm) \\
   \end{array}\right),
   \end{equation}
\begin{equation}\label{phi}
e^{i\phi_{\pm}}=\frac{\Delta(\gamma_\pm)}{|\Delta(\gamma_\pm)|}
\end{equation}
\begin{eqnarray}
 \left\{\begin{array}{cc}
\gamma_{+}=\arcsin{\frac{\hbar v_F q}{U_0+E_F}}\\
\gamma_{-}=\pi-\arcsin{\frac{\hbar v_F q}{U_0+E_F}}
\end{array}
    \right.
\end{eqnarray}
In the case of $d$-wave symmetry, the orientational dependency of
superconducting gap reads as
$\Delta(\gamma_{\pm})=\Delta(T)\cos(2\gamma_{\pm}-2\alpha)$ in which
$\alpha$ represents the orientation angle of $d$-wave
superconducting gap. We now proceed and using the above Dirac-BdG
wave functions and appropriate boundary conditions derive the
Andreev and Normal reflection coefficients. By applying appropriate
boundary conditions for the two interfaces which are located at
$x=0$ and $L$, we obtain all reflection and transmission
coefficients. At last we assume that a large gate voltage $V_0\gg 1$
is applied into the narrow ($L\ll 1$) normal region. In this case
$\Omega=V_0L/\hbar v_F$ is a constant which is called strength of
barrier. The approximation is called thin barrier approximation
regime in which the Normal region acts as an insulator. The Andreev
and Normal reflection coefficients in the thin approximation regime
are derived which are available in Appendix \ref{Andreev} for
F$\mid$I$\mid$$s$-wave superconductor junctions. We assume a
right-going electronlike quasiparticle within the ferromagnetic
region incident into interface between the ferromagnetic and
Insulator regions, so the appropriate boundary condition in the
interfaces at $x=0$ is:
\begin{eqnarray}\label{bncnl}
\nonumber &&\psi^{+}_{e,\sigma }(x)+ r_{A,\bar{\sigma}}
\psi^{-}_{h,\bar{\sigma} }(x)+r_{N,\sigma}
\psi^{-}_{e,\sigma }(x)=\\
&&t_{I,e}^{+}
\psi^{+}_{I,e}(x)+t_{I,e}^{-}\psi^{-}_{e}(x)+t_{I,h}^{+}\psi^{+}_{e}(x)+t_{I,h}^{-}\psi^{-}_{e}(x),
\end{eqnarray}
and other boundary condition in interface between the insulator and
superconductor regions at $x=L$ is:
\begin{eqnarray}\label{bncnr}
\nonumber &&t_{I,e}^{+} \psi^{+}_{I,e}(x)+t_{I,e}^{-}\psi^{-}_{e}(x)+t_{I,h}^{+}\psi^{+}_{e}(x)+t_{I,h}^{-}\psi^{-}_{e}(x)=\\
&&t_{S,e}^{+} \psi^{+}_{S,e}(x)+t_{S,h}^{-}\psi^{-}_{S,h}(x)
\end{eqnarray}
where $r_{A,\bar{\sigma}}$ and $r_{N,\sigma}$ are amplitudes of
spin-dependent Andreev and Normal reflection coefficients within the
ferromagnetic region, respectively. Other coefficients are
transmission coefficients in the Normal and superconducting regions.
By applying the thin barrier approximation on the obtained
reflection and transmission factors they reduce to simple factors
that are available in the Appendix \ref{Andreev}. For investigating
the electronic thermal conductance of the junction one needs to
calculate the probabilities of Andreev and Normal reflections namely
$|r_{A,\bar{\sigma}}|^{2}$ and $|r_{N,\sigma}|^{2}$. By assuming a
temperature gradient through the junction, the normalized thermal
conductance $\Gamma=\Gamma^{'}/\Gamma_0$ is given as
follow\cite{linder1,bardas}:
\begin{figure}
\includegraphics[width=7.5cm,height=6cm,clip]{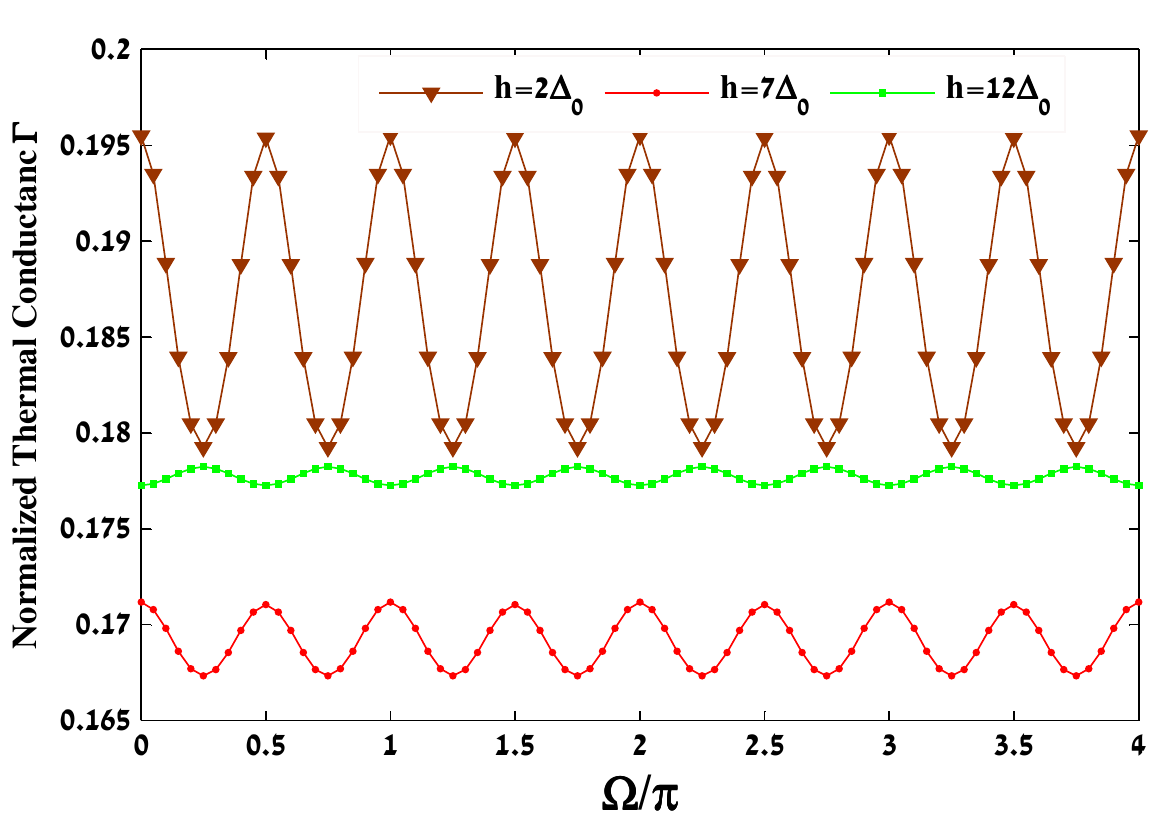}
\caption{\label{fig:KapavsOmegaswave} (Color online) The normalized
thermal conductance of F$\mid$I$\mid$S graphene-based junctions vs.
normalized strength of barrier region $\Omega/\pi$ for three values
of $h=2\Delta_0, 7\Delta_0, 12\Delta_0$. The temperature has been
fixed at $T=0.2T_c$. }
\end{figure}
\begin{eqnarray}
\nonumber\Gamma^{'}/\Gamma_0&=&\sum_{\sigma=\uparrow\downarrow}\int_{0}^{\infty}\int_{-\pi/2}^{\pi/2}d\epsilon
d\theta_{\sigma}\cos(\theta_{\sigma}) \{1-\mid
r_{N,\sigma}(\epsilon,\theta_{\sigma})\mid^{2}\\
&-&\mid
r_{A,\sigma}(\epsilon,\theta_{\sigma})\mid^{2}\}\frac{\epsilon^2}{T^2
\cosh^{2}(\frac{\epsilon}{2T})}, \label{Gamma}
\end{eqnarray}
where ${\Gamma_0}^{-1}=2\pi^{2}\hbar^{2}v_{F}k_{B}\Delta_0/E_F$ is a
constant. We proceed to investigate the characteristics of
electronic heat transport $\Gamma$ of the mentioned junctions and
throughout the paper we normalize energies with respect to
$\Delta_0$ and we set $\Delta_0=\hbar=k_B=1$ throughout our
computations.

\section{\label{FIS}Electronic thermal conductance of the F$\mid$I$\mid$S junctions in the thin barrier approximation }
\begin{figure}
\includegraphics[width=7.5cm, height=6.5cm]{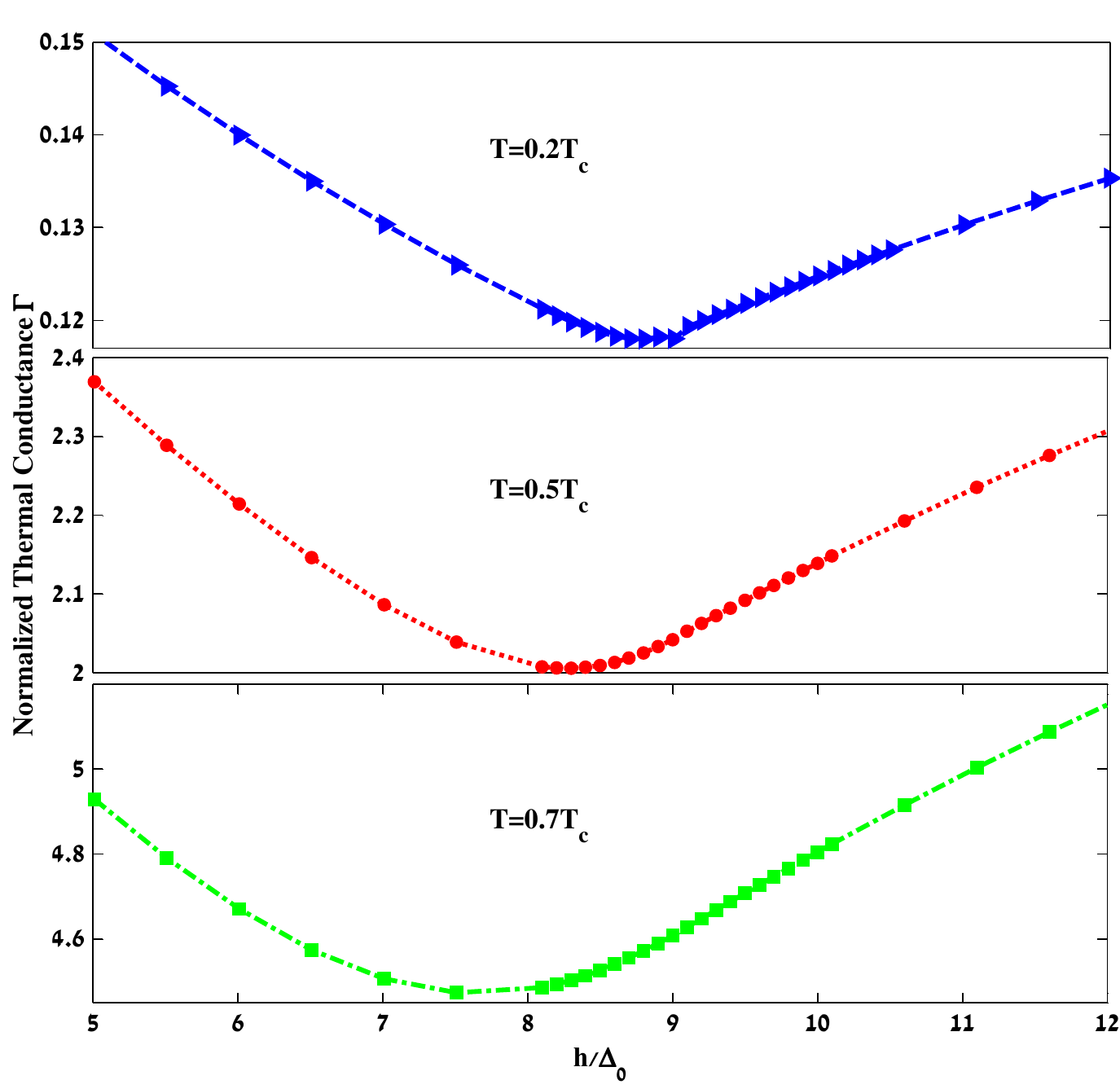}
\caption{\label{fig:Kapavshswave} (Color online) The normalized heat
conductance of F$\mid$I$\mid$S graphene-based junctions vs magnetic
exchange field strength $h/\Delta_0$ of Ferromagnetic region for
three values of temperatures, $T=0.2T_c, 0.5T_c, 0.7T_c$ and fixed
barrier strength at $\Omega=0$. }
\end{figure}

In this section we study electronic thermal transport
characteristics of the Ferromagnetic$\mid$Insulator$\mid$$s$-wave
superconductor junctions in the ballistic  and thin barrier
approximation regime. In Fig. \ref{fig:KapavsOmegaswave} we set
$T=0.2T_c$, and plot normalized thermal conductance $\Gamma$ vs
normalized strength of barrier $\Omega/\pi$ for three different
values of magnetization texture strength $h/\Delta_0$ , also in Fig.
\ref{fig:Kapavshswave} $\Omega=0$ is set and the normalized
conductance is plotted for three values of temperatures vs
magnetization texture strength $h/\Delta_0$. Throughout our
calculations we have set $E_F=10\Delta_0$ and also used a large
mismatch potential $U_0$. The normalized thermal conductance shows
an oscillatory behavior vs $\Omega/\pi$ which this finding can be
understood by noting the fact that how the amplitude of Andreev and
Normal reflections depend on $\Omega$, (See Appendix \ref{Andreev}).
In the thin barrier approximation, the width of normal layer $L$ and
barrier potential $V_0$ set for small and large values,
respectively. Andreev and Normal coefficients are involved $\cos
2\Omega$ and $\sin 2\Omega$ terms which are periodic functions of
$\Omega$ and consequently the appeared periodic oscillations in the
thermal conductance are originated from the two appeared periodic
functions in the Andreev and Normal coefficients. As it can be seen
in Fig. \ref{fig:model}, since the configuration contains one
semi-infinite superconductor the Andreev bound states don't
contribution to the transport characteristics of the junction under
consideration\cite{beenakker2,beenakker3,Sengupta,Sengupta2}. For
small values of magnetic strength, the amplitude of oscillations has
been enhanced in comparison with larger values of $h/\Delta_0$.

By increasing the magnetic strength, incident angle defined in Eq.
(\ref{incident angle}) reduce and hence diminish the amplitude of
oscillations which means suppression of available propagating
channels in the system. The normalized thermal conductance of the
F$\mid$I$\mid$S junction is plotted vs. $h/\Delta_0$, the
magnetization strength of Ferromagnetic region, in Fig.
\ref{fig:Kapavshswave} for three values of $T=0.2T_c, 0.5T_c,
0.7T_c$ and also $\Omega=0$ is set for the three plots. The thermal
conductance shows a minimum at $h\simeq E_F$ and by increasing
temperature move the minimum towards smaller values of $h$. The
magnetization texture splits Fermi level into two parts in the
$k$-space and by increasing $h$, the two parts separate upward and
downward more and more. Increasing the exchange splitting suppress
propagating Dirac Fermions modes in the configuration under
consideration up to values near $h\simeq E_F$, for larger values of
$h$ the propagating channels enhance , see Ref.s
\onlinecite{linder1} and \onlinecite{zareyan1}. In the light of
above discussion, the thermal conductance reach to its minimum value
at $h \simeq E_F$ that depends on the temperature. The fact also can
be inferred from Fig. \ref{fig:KapavsOmegaswave} which the curve of
$h=12\Delta_0$ has an intermediate value between the curves of
$h=2\Delta_0$ and $h=7\Delta_0$. We proceed to investigate effects
of $d$-wave symmetry on the heat conductance of F$\mid$I$\mid$d
junctions in the clean limit.

\begin{figure}
\scalebox{1.15}{\includegraphics[width=6.5cm, height=5.5cm]{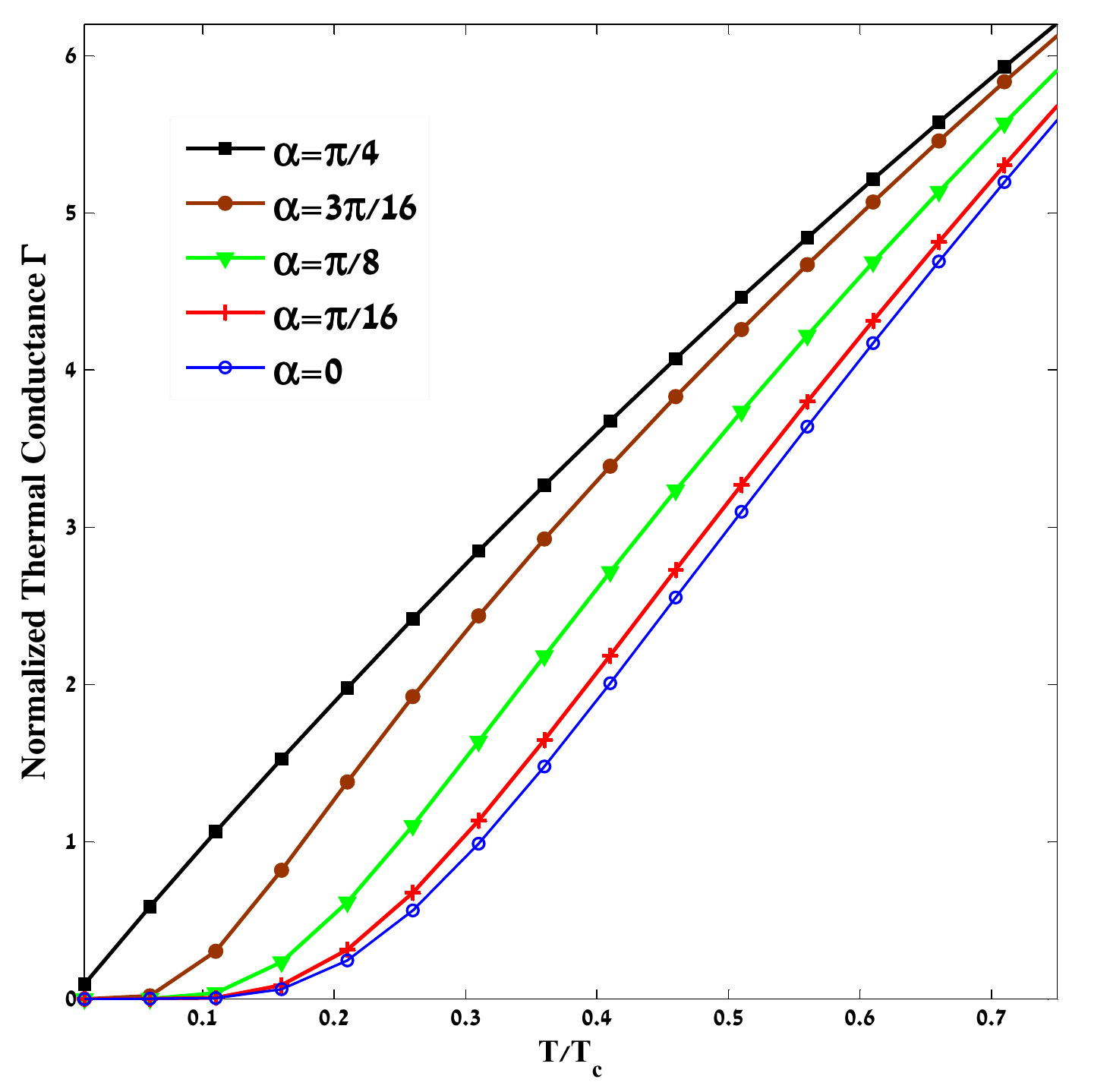}}
\caption{\label{fig:KapavsTdwave} (Color online) The normalized
thermal conductance $\Gamma$ of F$\mid$I$\mid$d graphene-based
junctions vs temperature for five values of $d$-wave superconducting
gap orientation $\alpha=0, \pi/16, \pi/8, 3\pi/16, \pi/4$. The
strength of magnetic exchange field and barrier fixed at
$h=2\Delta_0$, $\Omega=0$, respectively. }
\end{figure}
\section{\label{FId}Electronic thermal conductance of the F$\mid$I$\mid$d junctions in the thin barrier approximation }
Now we present main results of this paper namely the fingerprints of
$d_{x^2-y^2}$-wave superconducting region on the electronic heat
transport characteristics of F$\mid$I$\mid$d junctions in the thin
barrier regime whose interfaces are located at $x=0, L$. As it is
seen in Fig. \ref{fig:model}, unlike $s$-wave superconductors, the
role of crystal orientation of $d_{x^2-y^2}$-wave superconductors
with respect to interface is very important. We assume a
two-dimensional $d$-wave superconductor with cylindrical Fermi
surface in the $k$-space is deposited on top of graphene sheet and
connected to a sandwiched insulator region between Ferromagnetic and
superconducting regions. The pair potential for $s$-wave
superconductor is isotropic \textit{i.e.}
$\Delta(T)$=$\Delta_0\tanh\sqrt{1.76\sqrt{T_c/T-1}}$. On the other
hand, the pair potential for $d_{x^2-y^2}$-wave symmetry is
$\theta$-dependent, namely angle between the $a$-axis of the
superconductor crystal and wavevector of the conducting
quasiparticles. In this case, superconducting gap is anisotropic
\textit{i.e.} $\Delta_{\pm}(T,\gamma)$=$\Delta_{d}(T)\cos(2\gamma
\pm 2\alpha)$ in which $\alpha$ is angle of $a$-axis with respect to
normal trajectory to the interface (See Fig. \ref{fig:model}) and
$\gamma$ is propagation angle of quasiparticles. The temperature
dependency of $d$-wave superconductors is different from $s$-wave
case \cite{omelyanchuk}. As it mentioned above, for the thin barrier
approximation is assumed that $L\ll1$ and $V_0\gg1$, so one can
consider $\Omega$ as a constant and terms involving $\Omega$ reduce
to simpler ones. Here we have set $E_F=10\Delta_0$ and use large
mismatch potential $U_0$. Fig. \ref{fig:KapavsTdwave} indicates
electronic heat conductance of the F$\mid$I$\mid$d junctions vs
temperature for five different values of crystal orientation of
$d_{x^2-y^2}$-wave superconductor $\alpha$, the exchange field and
strength of thin barrier are set at $h=2\Delta_0$ and $\Omega=0$,
respectively. Electronic thermal conductance for $\alpha=0, \pi/16$
shows an exponential increase vs temperature and for larger values
of $\alpha$, the exponential form is modified to linear increase.
\textit{The heat conductance shows completely linear increase vs
temperature at maximum value of superconductor crystal orientation
angle $\alpha=\pi/4$ that induces in mind the Wiedemann-Franz law
for metals in low temperatures which thermal conductance is
proportional to temperature, $\Gamma \propto T$}.
\begin{figure}
\scalebox{0.9}{\includegraphics[width=9.5cm, height=8cm]{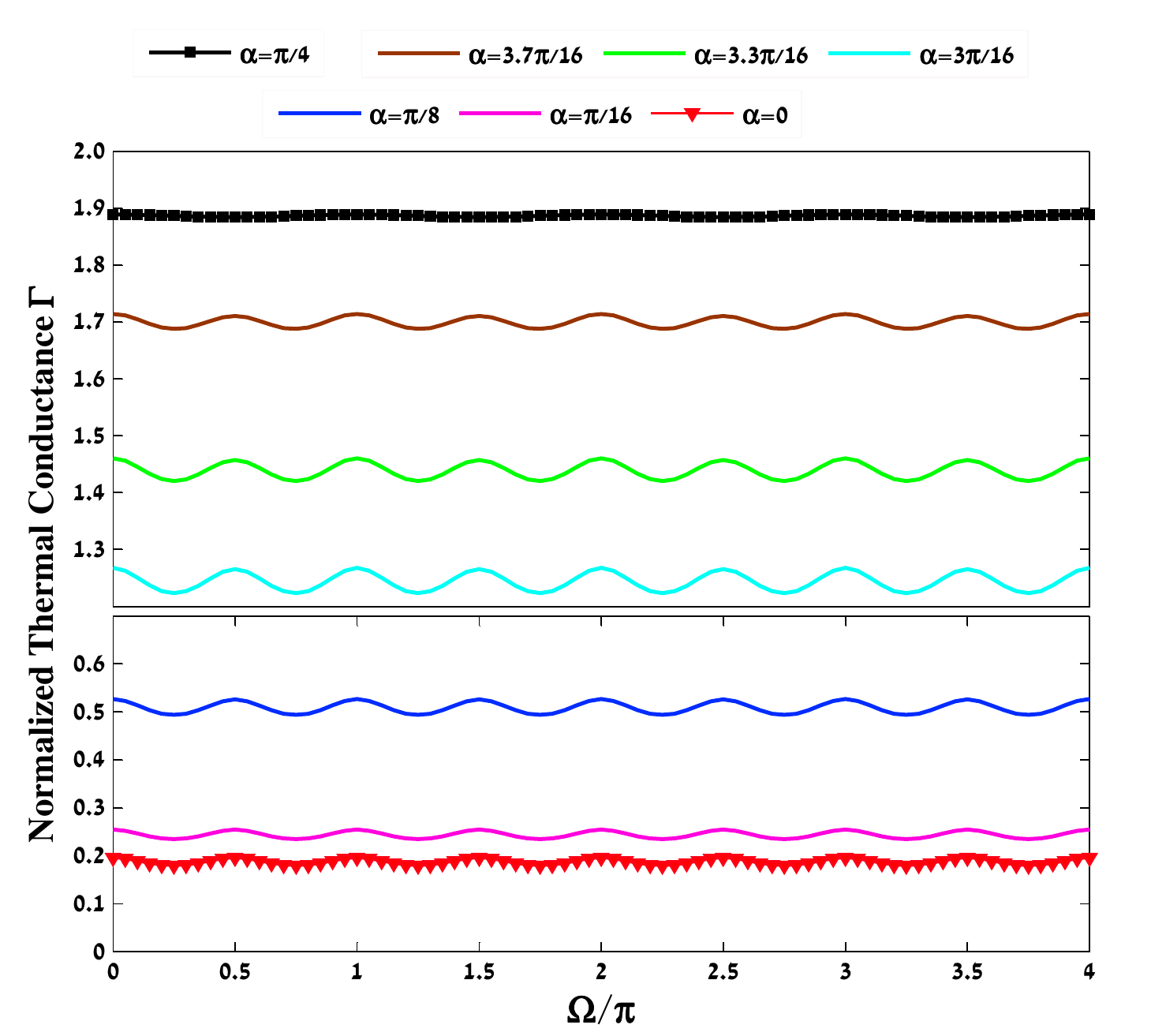}}
\caption{\label{fig:KapavsOmegadwave} (Color online) The normalized
thermal conductance $\Gamma$ of F$\mid$I$\mid$d graphene-based
junctions vs normalized strength of barrier region $\Omega$ for
seven values of superconducting gap orientation $\alpha=0, \pi/16,
\pi/8, 3\pi/16, 3.3\pi/16, 3.7\pi/16, \pi/4$. The strength of
exchange field and temperature are fixed at $h=2\Delta_0$,
$T=0.2T_c$, respectively. }
\end{figure}
The finding is arisen from orientational-dependent superconducting
gap that increasing $\alpha$ decreases the propagating channels of
superconducting correlations described by Andreev reflection
coefficients. Although the Dirac and Schrodinger equations are used
in graphene-based and metallic junctions respectively but the
behaviors of thermal conductance in the graphene-based junctions are
qualitatively similar to results of metallic N$\mid$I$\mid$d-wave
junctions in which the propagating channels of moving
quasi-particles is closed by increasing crystal orientation angle
from $0$ to $\pi/4$ in Ref.s \onlinecite{devyatov1} and
\onlinecite{devyatov2}. In this context $d$-wave symmetry shows the
same effects on thermal conductance of both graphene-based and
metallic junctions. The behaviors of heat conductance vs strength of
barrier are shown in Fig. \ref{fig:KapavsOmegadwave} for several
values of $\alpha$. Temperature and exchange field have set in
$h=2\Delta_0$ and $T=0.2T_c$. The thermal conductance vs strength of
barrier region shows an oscillatory behavior and the increase of
$\alpha$ enhance whole values of heat conductance.
\begin{figure}
\scalebox{0.95}{\includegraphics[width=7.5cm, height=8cm]{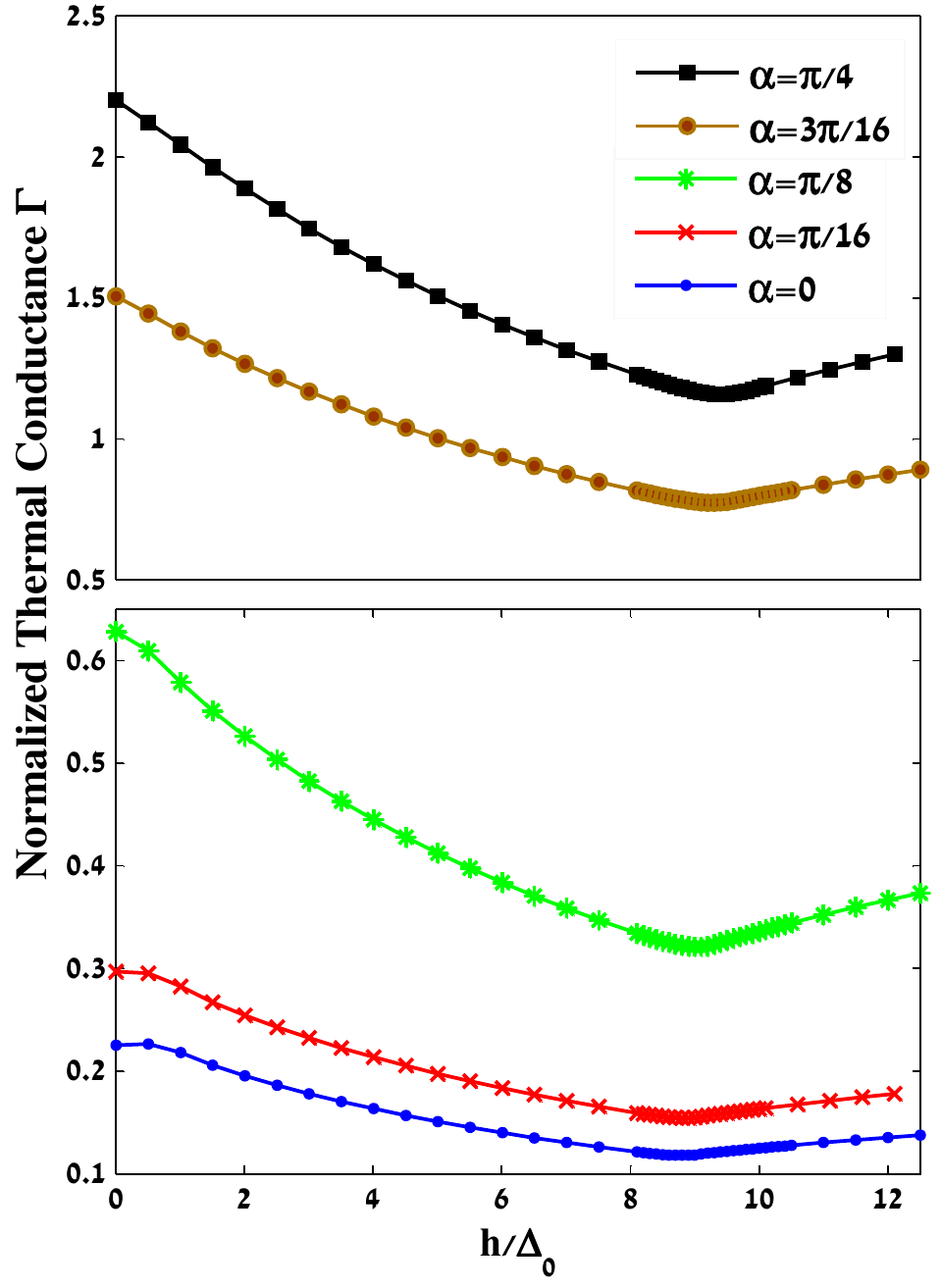}}
\caption{\label{fig:Kapavshdwave} (Color online) The normalized
thermal conductance of F$\mid$I$\mid$d graphene-based junctions vs
strength of magnetic exchange field $h/\Delta_0$ for five values of
superconducting gap orientation $\alpha=0, \pi/16, \pi/8, 3\pi/16,
\pi/4$. The temperature and strength of barrier are fixed at
$T=0.2T_c$, $\Omega=0$, respectively. }
\end{figure}
The period of oscillations vs strength of barrier suppresses
completely for maximum crystal orientation angle $\alpha=\pi/4$. In
Fig. \ref{fig:Kapavshdwave} the thermal conductance is plotted vs
the strength of magnetization exchange field $h/\Delta_0$ for
several values of $\alpha$ and $T=0.2T_c$. In general, $\Gamma$ for
F$\mid$I$\mid$d junctions vs $h/\Delta_0$ behaves similar to
F$\mid$I$\mid$S configuration . The behavior can be verified by
noting the mentioned reasons in the Sec. \ref{FIS} for
F$\mid$I$\mid$S case. Increasing the crystal orientation angle of
$d$-wave superconductor up to $\alpha=\pi/4$ can only enhance whole
values of the thermal conductance vs $h/\Delta_0$.

\section{Summary}\label{summary}
In summary we have considered
Ferromagnetic$\mid$Barrier$\mid$$s$/$d$-wave superconductors
graphene-Based junctions in the thin barrier approximation and
ballistic limit. We have utilized the Dirac-BdG equation and by
employing Dirac-BdG wavefunctions derived the Andreev and Normal
reflection amplitudes. Electronic thermal conductance $\Gamma$, of
the two mentioned junctions in the thin barrier approximation has
been investigated as well. We found that for F$\mid$I$\mid$S
junctions, the heat conductance vs magnetization strength
$h/\Delta_0$ shows a minimum at values near $h \simeq E_F$ that
 by increasing temperature the minimum move towards smaller values of
$h/\Delta_0$. The finding is qualitatively similar to
F$\mid$I$\mid$d junctions but increasing superconductive gap
orientation $\alpha$ shifts whole values of $\Gamma$ towards larger
values and no change induces to the trend of $\Gamma$ vs
$h/\Delta_0$. The electronic thermal conductance vs barrier strength
oscillates and shows identical behavior for F$\mid$I$\mid$S and
F$\mid$I$\mid$d configurations for all values of superconductor
crystal orientation $\alpha$ except values near $\alpha \simeq
\pi/4$. By approaching to $\alpha=\pi/4$, the propagating channels
diminish and hence the amplitude of oscillations suppress.
\textit{We found a Wiedemann-Franz law-like in the low temperature
regime for thermal conductance of F$\mid$I$\mid$S junctions, namely
$\Gamma \propto T$}. The electronic heat conductance shows an
exponential growth vs temperature for small values of gap
orientation angle $\alpha <\pi/8$ and for larger values especially
at $\alpha=\pi/4$ approaches to completely linear growth, namely
$\Gamma \propto T$.

\section*{Acknowledgments}
We thankful very useful and fruitful discussions with Jacob Linder.
The authors would like to thank the Office of Graduate Studies of
Isfahan University.
\appendix
\section{\label{Andreev}Andreev and Normal reflection coefficients in
the thin barrier approximation regime for F$\mid$I$\mid$S junctions}
Using the boundary condition Eq.s (\ref{bncnl}, \ref{bncnr}) and
applying the thin barrier approximation, the Andreev and Normal
reflection coefficients for the F$\mid$I$\mid$S junctions are
obtained as follows;
\begin{eqnarray}
\nonumber
\hspace{-1cm}r_A&=&\frac{\sqrt{\cos{\theta_{\sigma}}}\sqrt{\cos{\theta^{\prime}_{\bar{\sigma}}}}
\cos{\gamma}e^{(\frac{i(\theta_{\sigma}+\theta_{\bar{\sigma}})}{2})}}{\Upsilon_1+i\Upsilon_2}
\\
\nonumber \hspace{-1cm}r_N&=&\frac{e^{i\theta_{\sigma}}(\Sigma_1+i
\Sigma_2)}{\Upsilon_1+i\Upsilon_2}\label{rN}
\\
\nonumber
\Sigma_1&=&\cos{2\Omega}\cos{(\frac{\theta_{\sigma}+\theta^{\prime}_{\bar{\sigma}}}{2})}\sin{\beta}\sin{\gamma}
-\sin{(\frac{\theta_{\sigma}-\theta^{\prime}_{\bar{\sigma}}}{2})}\sin{\beta}
\\
\nonumber
\Sigma_2&=&\sin(\frac{\theta_{\sigma}-\theta^{\prime}_{\bar{\sigma}}}{2})\cos\beta
\cos\gamma+\sin2\Omega\cos(\frac{\theta_{\sigma}-\theta^{\prime}_{\bar{\sigma}}}{2})\sin\beta\sin\gamma
\\
\nonumber
\Upsilon_1&=&\cos{(\frac{\theta_{\sigma}-\theta^{\prime}_{\bar{\sigma}}}{2})}\cos{\beta}\cos{\gamma}+\sin{2\Omega}\sin{(\frac{\theta_{\sigma}+\theta^{\prime}_{\bar{\sigma}}}{2})}\sin{\beta}\sin{\gamma}
\\
\nonumber
\Upsilon_2&=&\cos{(\frac{\theta_{\sigma}-\theta^{\prime}_{\bar{\sigma}}}{2})}\sin{\beta}+
\cos{2\Omega}\sin{(\frac{\theta_{\sigma}-\theta^{\prime}_{\bar{\sigma}}}{2})}\sin{\beta}\sin{\gamma}.
\end{eqnarray}
The obtained coefficients recover the results of Ref.s
\onlinecite{beenakker1}, \onlinecite{zareyan1} and
\onlinecite{Sengupta} for N$\mid$S, F$\mid$S and N$\mid$I$\mid$S
graphene-based configurations, respectively. This can be justified
by letting $h\rightarrow 0$ and $\Omega\rightarrow 0$ in the above
coefficients for F$\mid$I$\mid$S graphene-based junctions.

\bibliographystyle{aip}

\end{document}